\documentclass{elsart1p}

\usepackage{graphicx}
\usepackage{amssymb}
\usepackage{amsmath,amssymb,amsfonts}
\usepackage{graphicx}
\usepackage{subfigure}
\usepackage{color}
\def \Slash{\slash \!\!\!}
\def \del{\partial}
\begin{document}
\begin{frontmatter}
%
%
%
%
%
\title{Quark Number Susceptibility and Thermodynamics
in HTL approximation}
%
%

 \author[sinp]{Najmul Haque\corauthref{cor1}},
 \ead{najmul.haque@saha.ac.in}
 \author[sinp]{Munshi G. Mustafa}
 \corauth[cor1]{Corresponding author}
 \address[sinp]{Theory Division, Saha Institute of Nuclear Physics,\\ 1/AF 
Bidhannagar, Kolkata 700064, India}

\begin{abstract}
In HTL perturbation theory we obtain leading order quark number 
susceptibility as a response to an external disturbance, {\em viz.},
chemical potential ($\mu$) that generates density fluctuation, which
is related to the correlation function through the thermodynamic sum 
rule associated with the symmetry of the system. We also obtain various 
thermodynamic quantities in leading order.

\end{abstract}

\end{frontmatter}

\section{Introduction}
\label{}
\vspace{-0.3cm}

The fluctuations of conserved charges such as baryon number related to
quark number susceptibility (QNS) is generally considered as a useful probe 
for quark gluon plasma (QGP). Various hard thermal loop (HTL) approaches and
methods \cite{blaizot,munshi,jiang} have led to different results for QNS.
We reconsider~\cite{haque} various thermodynamic quantities and QNS 
calculations by reformulating the HTL perturbation theory (HTLpt) to an 
external probe ($\mu$) and show that the leading order (LO) results agree
quite well despite the use of different approaches.  The QNS is defined 
through the thermodynamic sum rule as~\cite{haque,kunihiro} 
\vspace*{-.3cm}
\begin{eqnarray}
\chi(T) &=& \left.\frac{\partial \rho}{\partial \mu}\right |_{\mu=0}
= \left.\frac{\partial^2 {\cal P}}{\partial \mu^2}\right |_{\mu=0}
= \int d^4x \ \left \langle J_0(0,{\vec x})J_0(0,{\vec 0})
\right \rangle 
=-\lim_{p\rightarrow 0} {\mbox{Re}}\Pi_{00}(0,p),
\label{eq4}
\end{eqnarray}
where $\Pi_{\sigma\nu}(t,{\vec x})= 
\langle J_\sigma(t,{\vec x})J_\nu(0,{\vec 0})\rangle$ with external current 
$J_\sigma(t,{\vec x})$, ${\cal P}=(T/V)\ln{\cal Z}$ is the pressure,
 $\rho$ is the quark 
number density and ${\cal Z}$ is the partition function. For a given order in 
coupling $\alpha_s$, the thermodynamic consistency is automatic in (\ref{eq4}) 
for conventional perturbation theory whereas for resummed case one needs to 
take proper measure~\cite{haque}.

\section{HTL Perturbation Theory and Thermodynamic Sum Rule}
\vspace{-.2cm}
The HTL Lagrangian density~\cite{pisarski} for quark can be written as 
\vspace*{-.3cm}
\begin{eqnarray}
\mathcal{L}_{HTL} = \mathcal{L}_{QCD}+\delta{\mathcal{L}}_{HTL}
= \bar\psi i\gamma_\mu D^\mu \psi
+m_q^2\bar\psi\gamma_\mu\left
\langle \frac{R^\mu}{iR\cdot D}\right\rangle\psi \ , \label{i1}
\end{eqnarray}
where $R$ is a light like four-vector and $\langle\rangle$ is the 
angular average. A HTLpt has been developed~\cite{andersen} around 
$m_q^2$ by treating it as a parameter ($\sim (gT)^0$  and $(g\mu)^0$) 
much like a rest mass of a quark. In presence of an external source $\mu$, 
$D^\mu$ is  defined as 
${\tilde D}^\mu=({D}^\mu-i\delta^{\mu 0}\mu)$.
Expanding the second term around $\delta\mu$, we can write 
(\ref{i1}) as~\cite{haque}
\vspace*{-.23cm}
\begin{eqnarray}
\mathcal{L}_{HTL}(\mu+\delta\mu) 
&=& \bar\psi\left(i\Slash\!  {\tilde D} + \Sigma\right) \psi
+ \delta\mu \bar\psi \Gamma_0 \psi
+  \delta\mu^2\bar\psi \frac{\Gamma_{00}}{2}\psi
+ {\mathcal O}(\delta\mu^3) ,
\label{i2}
\end{eqnarray}
where $\Sigma,\Gamma_0 ,\Gamma_{00} $ are the various HTL $N$-point 
functions in coordinate space. The net quark number density (with 
$N_f$ flavour and $N_c$ color) 
becomes~\cite{haque}
\vspace*{-.23cm}
\begin{eqnarray}
\rho(\beta,\mu)\!\!=\!\! \left.\frac{\del{\cal P}}
{\del \mu}\right|_{\delta\mu=0}\!\! 
=\!N_cN_fT\!\! \int\!\! \frac{d^3k}{(2\pi)^3} 
\!\!\!
\sum_{k_0=(2n+1)\pi iT+\mu}
\!\!\!
\mbox{Tr} \left [S^\star(K)  \Gamma_0(K,-K) \right].  
\label{i3_3} 
\end{eqnarray} 
Similarly, the QNS becomes~\cite{haque}
\begin{eqnarray}
\chi(\beta)\!&=&\!\!
\left.\frac{\del{\rho}}{\del \mu}
\right|_{\mu = 0}\!\!\!
=\left.\frac{\del^2{\cal P}}{\del \mu^2}
\right|_{\mu = 0} 
= - N_cN_fT\!\! \int \frac{d^3k}{(2\pi)^3}
\sum_{k_0=(2n+1)\pi iT}
 \nonumber \\  
&\!\!\!\times&
\mbox{Tr} \left [S^\star (K)  \Gamma_0(K,-K;0)S^\star(-K)
\Gamma_0(K,-K;0)
 - S^\star (K) \Gamma_{00}(K,-K;0,0)\right ].\ \ \ \ \ 
\label{s1}
\end{eqnarray}
The first term in (\ref{s1}) is a one-loop self-energy
whereas the second term is a tadpole in HTLpt with effective
$N$-point functions at the external momentum $P=(\omega_p,p)=0$. This 
is thermodynamic sum rule for LO HTLpt associated with quark number 
conservation (gauge symmetry). As we will see below it does not matter 
which of the equivalent expression in (\ref{s1}) is used to
obtain QNS in contrast to Ref.~\cite{jiang}. 

\section{Thermodynamics and Quark Number Susceptibility}
\vspace{-.3cm}
We intend first to obtain the LO number density $\rho(T,\mu)$ from 
(\ref{i3_3}) and then various thermodynamic quantities and QNS. 
Now considering only first term of (\ref{i3_3}), 
the quasi-particle (QP) part of the LO quark number density (with 
subscript $I$)  becomes~\cite{haque}  
\begin{eqnarray}
\rho_I^{QP}(T,\mu)&=&2N_cN_f\int \frac{d^3k} {(2\pi)^3}\left[
n(\omega_+-\mu)+n(\omega_--\mu)-n(k-\mu) -\{\mu\rightarrow-\mu\}
\right ],   \ \ 
\label{H6}
\end{eqnarray}
which agrees with that of Ref.~\cite{blaizot}. Now, the Landau damping (LD) 
part~\cite{haque} reads as 
\vspace{-.3cm}
\begin{eqnarray}
\rho_I^{LD}(T,\mu)= N_cN_f\int \frac{d^3k}{(2\pi)^3}\int \limits_{-k}^k 
{d\omega}
\left(\frac{2m_q^2}{\omega^2-k^2}\right)\beta_+(\omega,k)
\left[n(\omega-\mu)- n(\omega+\mu) \right ]  \ ,
\label{L3}
\end{eqnarray} 
where $n(y)$ is the FD function and $\beta_\pm$ is the cut-spectral 
function of HTL propagator~\cite{haque}.
From this number density (QP+LD) expression, pressure and entropy 
density can be obtained~\cite{haque} using thermodynamics relations and
they agree with those of Refs.~\cite{blaizot,andersen}.

\begin{figure}[!tbh]
\subfigure{
{\includegraphics[height=0.27\textwidth, width=0.45\textwidth]{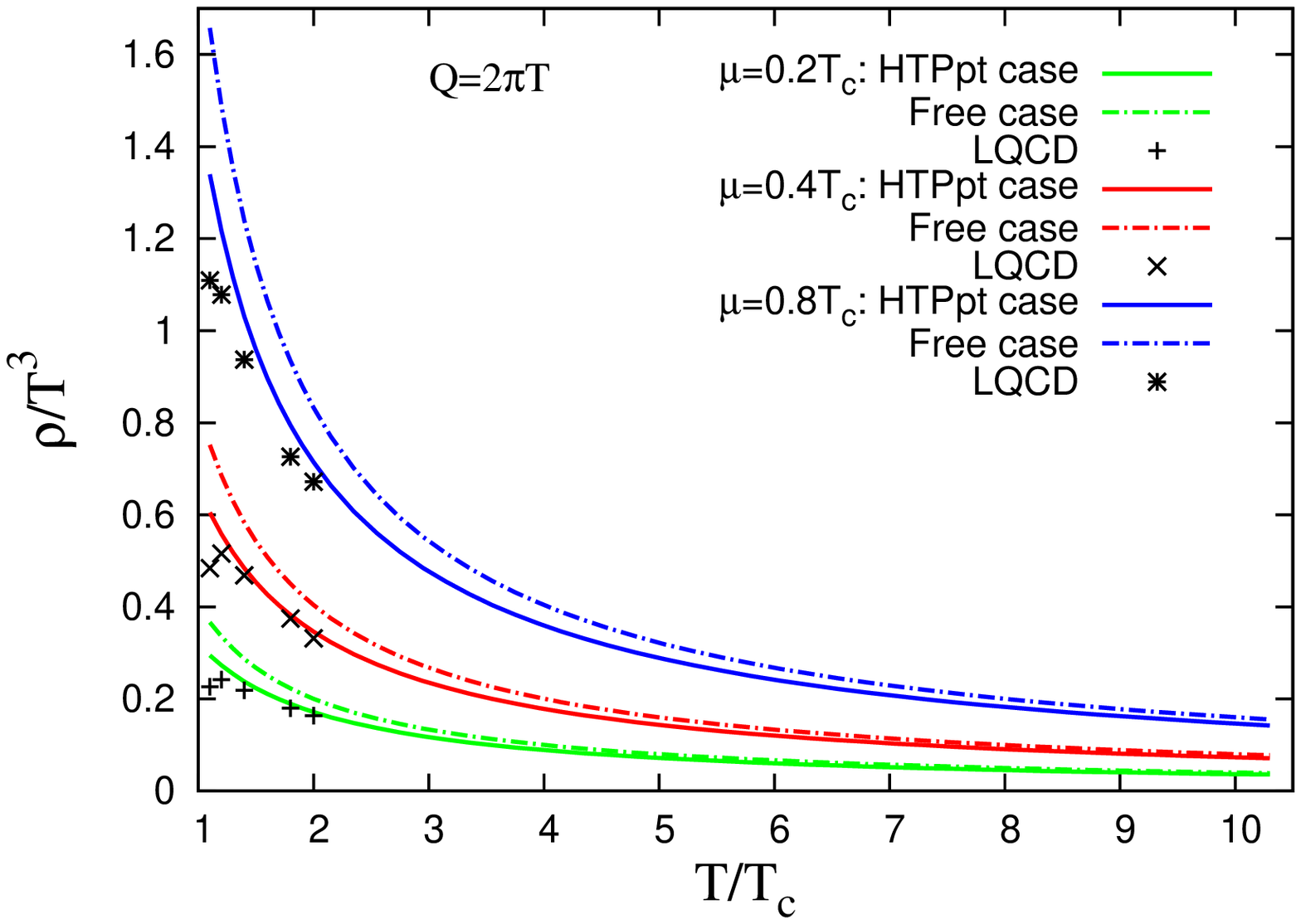}}
}
\subfigure{
{\includegraphics[height=0.27\textwidth, width=0.45\textwidth]{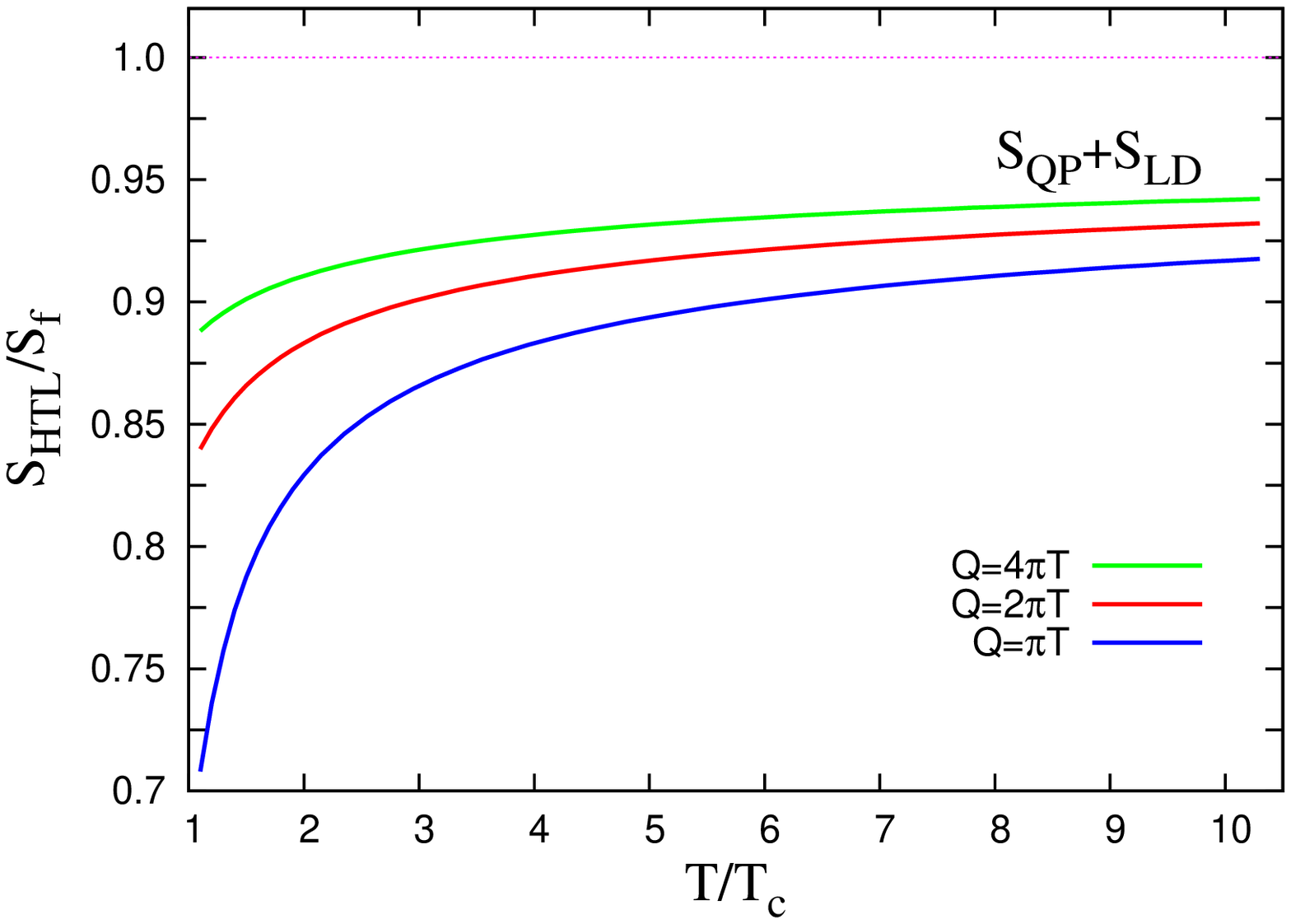}}
}
\vspace*{-0.2in}
\caption{(Color online){\em{Left panel}}: Scaled LO quark number
 density with $T^3$ vs $T/T_C$ for various $\mu/T_C$. Lattice data 
are from~\cite{alton}.{\em{Right panel}}: Scaled entropy density 
with free one vs $T/T_C$ for $\mu=0.2T_c$. $T$ dependent $\alpha_s$
is used with scal $Q$~\cite{haque}.} 
\label{rho_I}
\end{figure}

The LO  QNS is obtained~\cite{haque} by taking the derivative of $\rho_I$ 
w.r.t $\mu$  as
\begin{eqnarray}
\!\!\!\!\!\!
\!\!\!\!\!\!
\chi_I^{QP}(T)=4N_cN_f\beta\!\!\!\int\!\!\!\frac{d^3k} {(2\pi)^3}
\left [ n(\omega_+)\left(1-n(\omega_+)\right) 
 + n(\omega_-)\left(1-n(\omega_-)\right)- n(k)\left(1-n(k)\right)
\right ],\nonumber   
\end{eqnarray}
\vspace{-.8cm}
\begin{eqnarray}
\chi_I^{LD}(T)
=2N_cN_f\beta\int\frac{d^3k}{(2\pi)^3} 
\int\limits_{-k}^k d\omega\left(\frac{2m_q^2}{\omega^2-k^2}\right) 
\beta_+(\omega,k)\ n(\omega)\left(1-n(\omega)\right) \ , \label{L4}
\end{eqnarray}
where the $\mu$ derivative is performed only to the explicit $\mu$
dependence and agrees with Ref.~\cite{blaizot}. The LO HTL 
quark density and entropy 
density for $2$ flavour as a function of $T/T_c$ are shown in 
Fig.~\ref{rho_I} whereas HTL QNS is shown in the 
left panel of Fig.\ref{pert_g2}.

\vspace{-.6cm}
\section{Correlation Function, Thermodynamic Sum Rule and QNS}
\vspace{-.4cm}
Now, we  obtain QNS (denoted by subscript $II$) from the correlation 
function of (\ref{s1}) that relates the thermodynamic sum rule 
associated with the conserved baryon number.
The QP  and LD part of QNS are obtained~\cite{haque}, respectively, as
\begin{eqnarray}
\chi_{II}^{QP}(T)&=& 4N_cN_f\beta\int \frac{d^3k}{(2\pi)^3}
\left. \Big [n(\omega_+)\left
(1-n(\omega_+)\right)+n(\omega_-)\left(1-n(\omega_-)\right)\right.\nonumber \\
&& \left.   \ \ \ \ \ \ \ \ \ \ \ \ \ \ \ \ \ \ \ \ \ \ \ \ \ 
-n(k)\left(1-n(k)\right) - \frac{1}{\beta k}(1-2n(k)) \right],
\nonumber \\ 
\chi_{II}^{LD}(T)
&=&2N_cN_f\beta\int\frac{d^3k}{(2\pi)^3}\int\limits_{-k}^k d\omega
\left(\frac{2m_q^2}{\omega^2-k^2}\right)\ \beta_+(\omega,k)\ n(\omega)
\left(1-n(\omega)\right)\ \nonumber \\
&&\ + N_cN_f\int\frac{d^3k}{(2\pi)^3}\int\limits_{-k}^k d\omega
\left(\frac{2m_q^2}{\omega^2-k^2}\right)\ \frac{\beta_+(\omega,k)}{\omega+k}
(1- 2n(\omega)) \ . \label{s15}
\end{eqnarray}
Both expressions in (\ref{s15}) agree to those, respectively, in (\ref{L4})
but with an additional term. For LO QNS one can neglect~\cite{haque} these 
additional terms in (\ref{s15}). Thus, the LO thermodynamic sum rule 
associated with quark number conservation in HTLpt 
is guaranteed. Below we show the correct inclusion of LO to QNS in HTLpt 
in a {\em strict perturbative sense}.

\vspace{-.2cm}
\section{QNS in Perturbative Leading Order ($g^2$)}
\vspace{-.4cm}
In conventional perturbation theory, the QNS
has been calculated for massless QCD~\cite{blaizot} upto 
order $g^4\log(1/g)$ at $\mu=0$ as  
\begin{equation}
\frac{\chi_p}{\chi_f}=1- \frac{1}{2}\left (\frac{g}{\pi}\right )^2 +
\sqrt{1+\frac{N_f}{6}}\left(\frac{g}{\pi}\right )^3-
\frac{3}{4}\left(\frac{g}{\pi}\right )^4 \log\left(\frac{1}{g}\right)+
{\cal O}(g^4)\ . \label{qns_pert}
\end{equation}

\begin{figure}[!tbh]
\subfigure{
{\includegraphics[height=0.28\textwidth, width=0.312\textwidth]{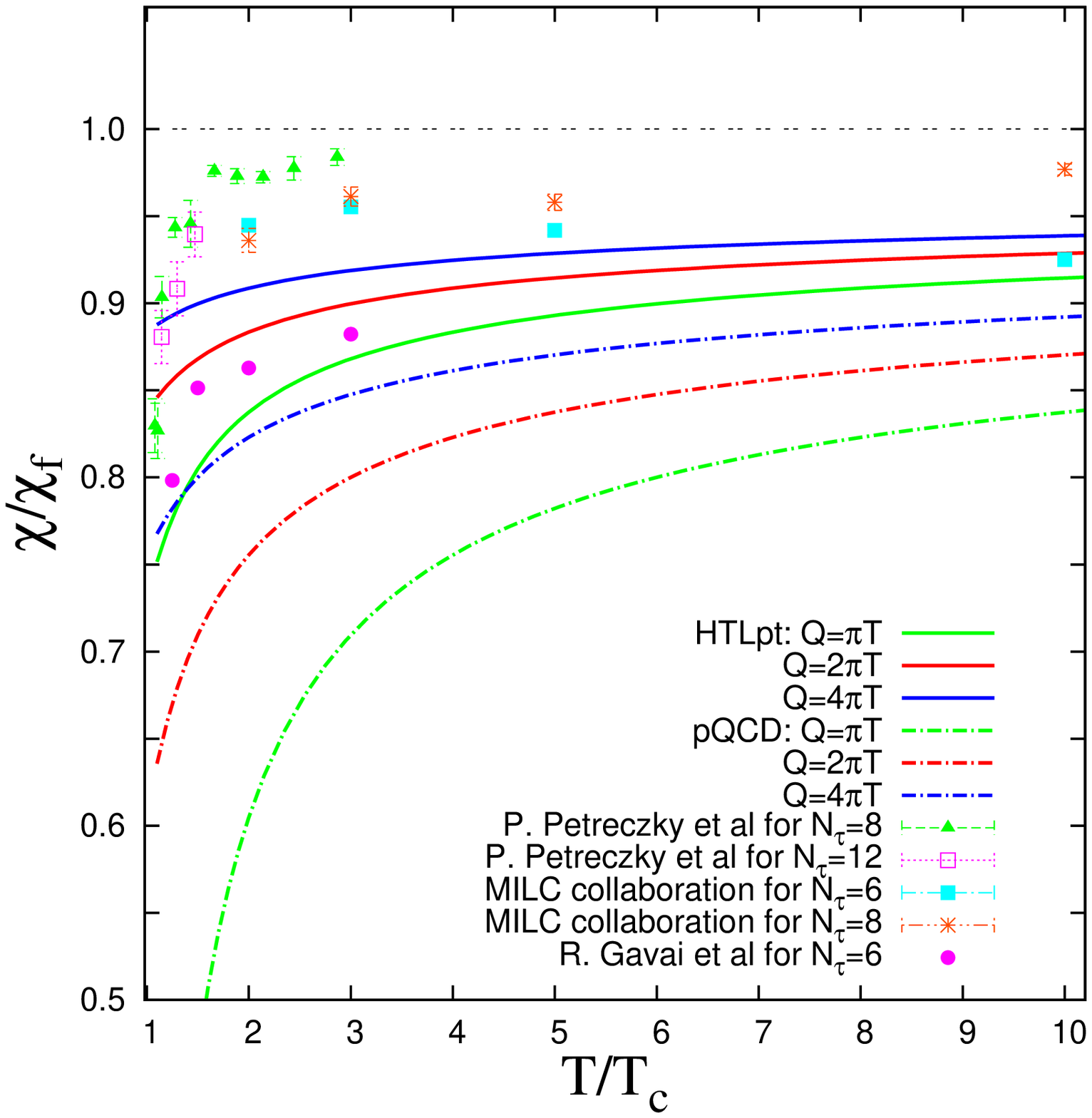}}
}
\subfigure{
{\includegraphics[height=0.28\textwidth, width=0.31\textwidth]{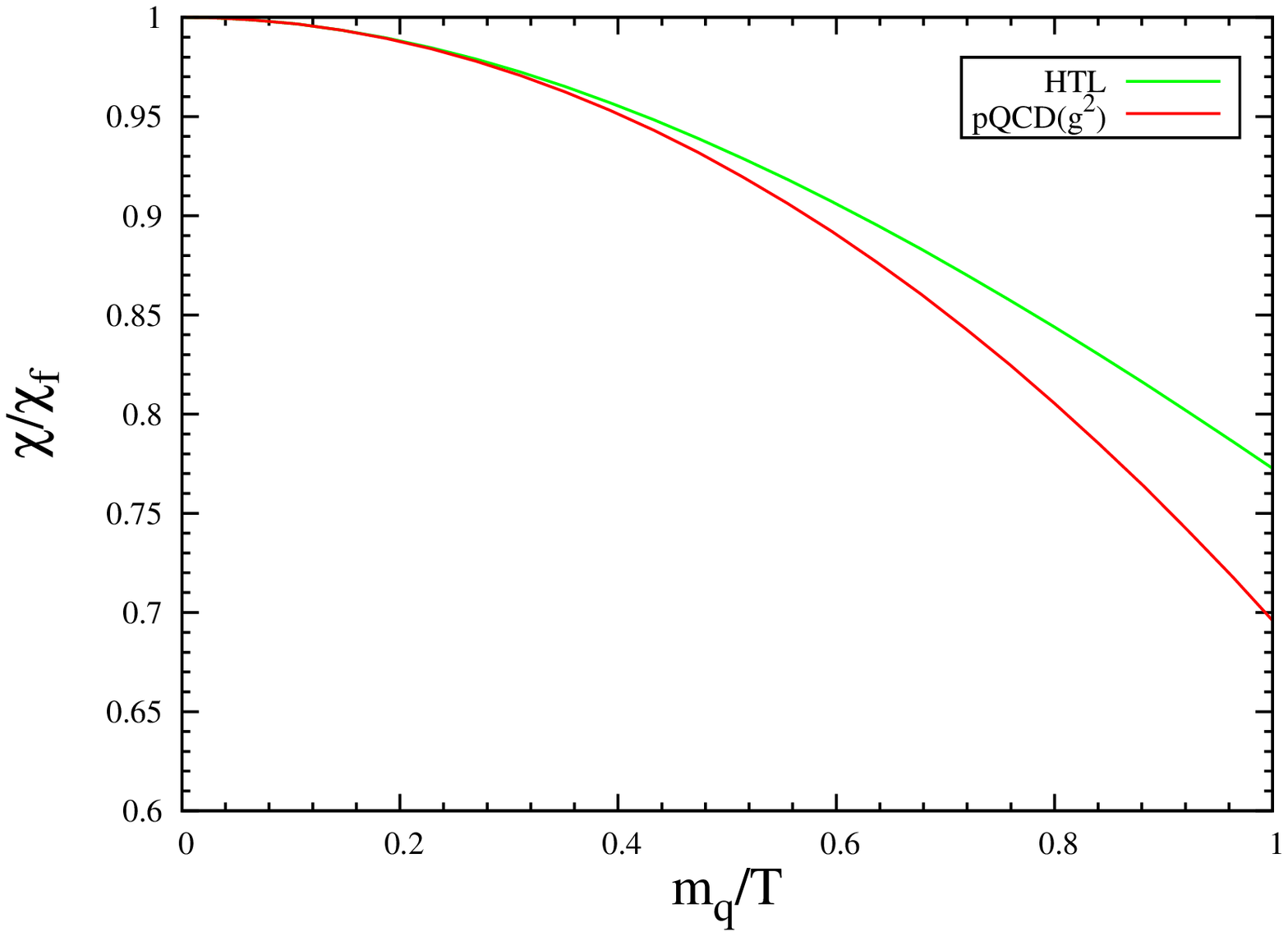}}
}
\subfigure{
{\includegraphics[height=0.28\textwidth, width=0.31\textwidth]{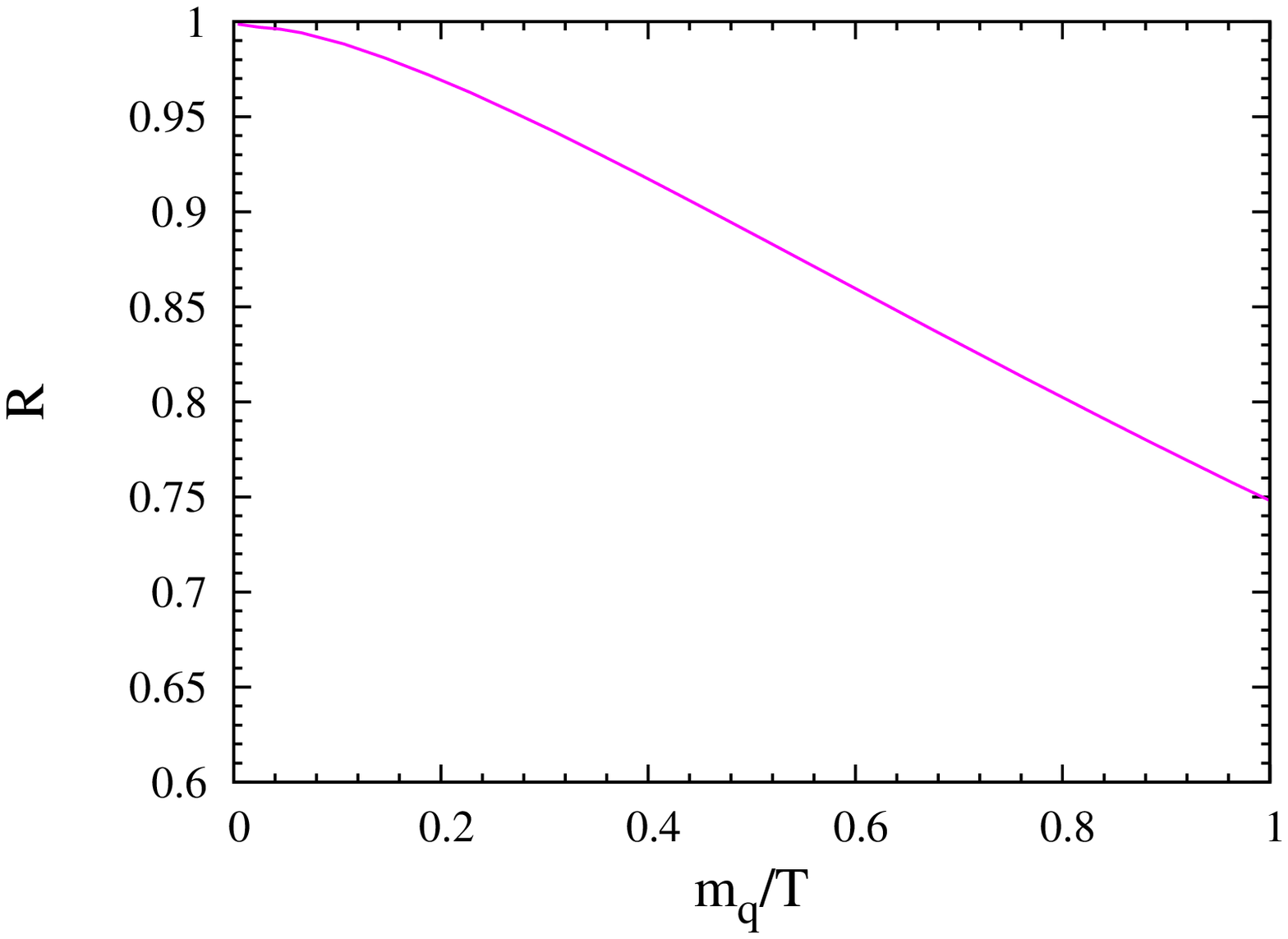}}
}
\vspace*{-0.22in}
\caption{(Color online) {\textit {Left panel:}}
The 3-flavour scaled QNS with that of free one as a function of $T/T_C$. 
Symbols represent various lattice data~\cite{lat2}.
{\textit {Middle panel:}} 
The ratio of 2-flavour HTL to free quark
QNS and also that of LO  perturbative one  as a 
function of $m_q/T$. {\textit {Right panel:}} 
$R$ as a function of $m_q/T$.}
\label{pert_g2}
\end{figure}

In the middle panel of Fig.~\ref{pert_g2} we display the LO HTL and
perturbative QNS scaled with free one vs $m_q/T$. In the weak 
coupling limit both approach unity whereas HTL has a little slower deviation 
from ideal gas value. Now, in the right panel we plot a ratio~\cite{blaizot},
$R\equiv ({\chi_{htl}-\chi_f})/({\chi_{p(g^2)}-\chi_f})$, 
which measures the deviation of interaction of $\chi_{htl}$ from that of pQCD
in order $g^2$. It approaches unity in the weak coupling limit
implying the correct inclusion~\cite{blaizot}
of order $g^2$ in our approach in a truly perturbative sense. 

\section{Conclusion}
\vspace{-0.2cm}
We have formulated a thermodynamically consistent HTLpt at the first 
derivative level of the the thermodynamic potential by incorporating 
an external source, {\em viz.}, the quark chemical potential to the system. 
We show that the various thermodynamic quantities and the QNS agree with 
the other HTL approaches~\cite{blaizot} existing in the literature.
In addition we also obtained the QNS from the correlation functions
associated with the conserved number density fluctuation of the system. 
The equivalence of the results obtained for the QNS in LO HTLpt in two ways 
(Sec. 3 and 4)gurantees the thermodynamic sum rule, which was not shown  
earlier within HTL approximation earlier. We also reproduce the LO 
perturbative results in the weak coupling limit, which indicates the correct 
inclusion of the LO in a strict perturbative sense in our approach.

\end{document}